\documentclass[11pt,reqno]{amsart}

\usepackage{amsmath}
\usepackage{amssymb}
\usepackage{graphicx}
\usepackage{amscd} 
\usepackage{stmaryrd}
\usepackage[mathscr]{euscript}

\newtheorem{theorem}{Theorem}[section]

\theoremstyle{definition}

\numberwithin{equation}{section}

\title[Commuting Integral and Differential Operators]{Commuting Integral and Differential Operators and the master symmetries of the Korteweg-\lowercase{de} Vries equation}

\author[F. Alberto Gr\"unbaum]{F. Alberto Gr\"unbaum}
\address{
Dept of Mathematics \\
University of California, Berkeley \\
Berkeley, CA 94720 \\
U.S.A.
}
\email{grunbaum@math.berkeley.edu}



\date{\today}

\begin{document}

\begin{abstract}

   The singular value decomposition going with many problems in medical imaging,
   non-destructive testing, geophysics, etc. is of central importance.
    Unfortunately the effective numerical
   determination of the singular functions in question, is a very ill-posed problem.
   The best known remedy to this problem goes back to the work of D. Slepian, H. Landau
   and H. Pollak, Bell Labs 1960-1965.

We show that the master symmetries of the Korteweg-de Vries equation give a way
to extend the remarkable result of D. Slepian in connection with the Bessel
integral kernel and the existence of a differential operator that commutes with  the corresponding integral operator. The original results of Bell Labs group has already played an important role in the study of the limited angle problem in X-ray tomography as well as in Random matrix theory.

\end{abstract}

\maketitle






\section{Introduction}

   This section starts with a reader friendly introduction to the two themes
   mentioned in the 
   title and the relevance of the first one to an important image reconstruction
   problem. This is followed by a more detailed introduction and a description of the new results in the paper.

\subsection{The work of the Bell Labs group, commuting integral and differential operators}

   Starting in 1961, the Bell Labs group of H. Landau, H. Pollak and
   D. Slepian produced a series of remarkable papers under the general
   title "Prolate spheroidal wave functions, Fourier analysis and uncertainty"
   \cite {SP,LP1,LP2,S,SV}.
   The first few of these papers, papers I,II and III, deal with the analysis of signals in
   continuous time. Paper IV in the series is a paper by D. Slepian 
   and it considers the multidimensional case. His results 
   will be extended in the present paper, as will be explained later in this section.

   The last paper in the series, paper V, \cite{SV},
   is once again by D. Slepian and it considers the case
   when the time series is given by discrete samples of a signal. This
   naturally leads him to consider the amplitude spectrum given by a function
   defined on the unit circle, i.e. one replaces the Fourier transform
   in one or several variables by the Fourier series.

   When one looks at these different scenarios for a unified signal processing
   point of view, the different situations deal with various incarnations of
   the same problem which originated with C. Shannon: what is the best use
   one can make of a limited duration portion of a signal if one knows that
   the signal is bandlimited, i.e. its Fourier content vanishes outside
   of an interval on the real line, a disc in ${R}^{N}$ or, in the last case, an
   interval on the unit circle?

   The only setup of the Fourier picture not
   considered by the Bell Labs group is the purely finite one, namely 
   the case when both physical and frequency space is the set
   of N-roots of unity, i.e. the Discrete Fourier Transform, which was later
   considered in \cite{G0}.

   Returning to paper V in the series, see \cite{SV},  involving discrete
   time and the Fourier  series, D. Slepian looks at the appropriate Toeplitz matrix in section 2 of the
   paper and establishes a series of asymptotic results for its eigenvectors
   and eigenvalues, see in particular section 2.5. His Toeplitz matrix has 
   entries given by $$\frac{2 \phi}{\pi}$$ on the main diagonal and by
   $$\frac{2 \sin (l \phi)}{\pi l}$$ for the $l$-th diagonal.

\bigskip

   After section 2 in \cite{SV} one finds a long section on applications of the analytical
   results to problems in signal processing. I reproduce the list of applicatons discussed by D. Slepian to indicate its wide scope.
    In section 3.1 one finds: discussions of  extremal properties,
    most concentrated bandlimited sequence,
    index limited sequence with most concentrated spectrum,
   simultaneously achievable concentration,
    minimum energy bandlimited extension of a finite sequence,
    trigonometric polynomial with greatest fractional energy in an 
          interval-optimal window.
Section 3.2 deals with  
   a prediction problem, and finally section 3.2 deals with
    the approximate dimension of signal space.

\bigskip

\bigskip

   It is remarkable that the problem considered by D. Slepian has (at least) one more
   application, maybe closer to the interests of workers in Inverse Problems.

   In the book The Mathematics of Computerized Tomography,\cite{N} section VI.2, entitled ``the limited angle problem",
   F. Natterer gives a full discussion of the relevance of D. Slepian's work
   to analyze the degree of ill-conditioning of the reconstruction problem when
   only projections limited to the angle $\phi <\pi/2$ are considered.
   The matrix that is analyzed in \cite{N} is exactly the same as in \cite{SV}
   and the main result quoted from this reference in page 161 of \cite{N}, 
   shows in a sharp quantitative fashion that when the size of the matrix
   is large a small fraction of its eigenvalues are close to $1$ and most 
   of the eigenvalues are very close to $0$ with a very small transition region in
   between. The fraction in question depends on the angle $\phi$ in ``the limited angle problem".
   For a fuller discussion of the limited angle problem the reader should consult \cite{N} as well as \cite{G1}. 

   Needless to say the fact that most of the eigenvalues are very
   close to $0$ makes the computation of the corresponding eigenvectors a
   difficult problem, which is handled, as in all the other papers of the Bell Labs group
    by the existence of a second order differential (or a tridiagonal
   matrix in paper V) that commutes with the original convolution integral operator (or
   a Toeplitz matrix in the case of paper V).

\bigskip

The Bell Labs group found a mathematical miracle: one
   can produce a differential operator with simple spectrum that shares its eigenfunctions with the integral operator in question and
   has a very spread out spectrum, resulting in a numerically stable problem when it comes to computing these eigenfunctions.

\bigskip

   The discussion above
   illustrates very well the fact that long series of papers mentioned
   earlier, see \cite{SP,LP1,LP2,S,SV} have found important applications in areas far
   removed from the
   original problem considered in them. A few more instances of this are seen
   in \cite{SD,SDW,TW3}.

\bigskip

   In 1982 D. Slepian delivered the John von Neumann lecture at the SIAM meeting,
   see \cite{S1}. After giving a short view of the work done by the Bell Labs
   group he closes the introduction with ``The mystery of this serendipity grows.
   Most of us feel that there is something deeper here that we currently
   understand-that there is a way of viewing these problems more abstractly that
   will explain their elegant solution in a more natural and profound way, so 
   that these nice results will not appear so much as a lucky accident". 

   The new results given in this paper, to be described in Section $5$,
   are a extension of the results in paper IV of the series of papers
   discussed above and could be considered in the spirit of the comments above, even if 
   no one has found a concrete application of them yet. Presenting them to an
   audience of people interested in applications should increase the chance
   that an application may be found.

\subsection{The Korteweg-de Vries equation}

   Starting around 1967 people started finding ways to produce explicit
   solutions of a non linear partial differential equation in the theory of water waves,
   namely
$$u_t = u_{xxx}- 6 u u_x.$$

  Solitary wave solutions had been found much earlier but the discovery
  of ``soliton type solutions" opened up a remarkable period of interesting
  developments. This is not the place to take even a brief look at this
  area, but the interested reader may want to look (for instance) at \cite{FZ}.

\bigskip

  A look at history shows that the development of methods to solve
  the heat and the wave equations 
  produced tools that went way beyond the original
  problem. One can only hope that some of the developments centered around the Korteweg-de Vries equation (KdV equation from now on) could
  have payoffs in areas very far from their birth place. The execution of this
  program is a challenge.

  One way to make the statement above less controversial is to notice that the
  KdV equation can be seen as the second in a hierarchy of partial differential
  equations of which the first one is given by

$$u_t=u_x.$$

  This linear partial differential equation is trivial to solve and its solution is given by the
  time translation of the initial data.
  The point is that translation is the underlying structure behind Fourier
  analysis.

\subsection{Statement of the new results}

The operator
\begin{equation}
\label{Bessel}
L_\nu = -D_x^2 + \frac {\nu^2-1/4}{x^2}\qquad x > 0
\end{equation}

\noindent
plays a crucial role in mathematical physics, geometry and many other areas. The reason behind this is very simple: after conjugation it gives the radial part of the (negative) Laplacian in $R^N$ when $\nu= \frac{N-2}{2}$.

The space of its eigenfunctions, i.e., solutions of
\[
L_\nu\varphi(x,z) = z^2\varphi(x,z)
\]
is given in terms of the well known Bessel functions and the bounded solution at $x = 0$ is
\begin{equation}
\label{eigenf}
f_\nu(x,z) = \sqrt{xz} J_\nu(xz).
\end{equation}

In this paper we will revisit a remarkable property of $L_\nu$ discovered by D.~Slepian a long time ago \cite{S}.
This has important ``signal processing'' applications. His result is an extension to the Bessel case of a property of the Fourier transform in $R$ which had been studied by the Bell Labs group \cite{SP,LP1}. In the Fourier case (obtained by setting $\nu =-\frac{1}{2}$) they showed that the integral operator with kernel

\[
K_T(z_1,z_2) = \int_{-T}^T e^{iz_1x}e^{-iz_2x} dx = \frac {\sin T(z_1-z_2)}{z_1-z_2}
\]
acting on $L^2([-G,G],dz)$ commutes with the differential operator
\[
-D_z(G^2-z^2)D_z + z^2T^2.
\]
D.~Slepian found that if one considers the eigenfunction of $L_\nu$ given by $f_\nu(x,z) = \sqrt{x z}
J_\nu(xz)$ then the integral operator with kernel

\begin{equation}
\label{kernel}
K(z_1,z_2) \equiv \int_0^T f_\nu(x,z_1)f_\nu(x,z_2)dx
\end{equation}
acting in $L^2((-G,G),dz)$ admits a commuting differential operator, namely
\begin{equation}
\label{commut}
	\mathbb{A}_\nu= -D_z(G^2-z^2)D_z + z^2T^2 + G^2\frac {\nu^2-1/4}{z^2}\,.
\end{equation}

\bigskip

The eigenfunctions of the integral operators with kernel \eqref{kernel}  are important in signal processing since they are the {\bf singular functions} of important {\bf time-and-band limiting} problems.

\bigskip

  It was metioned earlier in this introduction that
  in the context of  inverse problems such as X-ray tomography these
  ideas have played am important role, specifically in the ``limited-angle
  problem", see chapter $6$ of F. Natterer's book, \cite{N} section VI.2, where one can see
  the connections with the work of D.Slepian, \cite{SV}. For a vey good
  discussion of the general problem in signal processing, see \cite{S1,S2}.

\bigskip

  In the setup of limited angle tomography ``time limiting'' is replaced by the fact that the function that one wants to reconstruct (a slice of the patient's body) has compact support while the knowledge of the line integrals orthogonal
  to directions in a limited angle amount the knowledge of the Fourier transform of the object in this limited angle because of the ``central section theorem". This plays the role of ``band limiting". For a more complete discussion connecting time and band limiting to the integral operators above in the context of X-ray tomography see \cite{G7}.

  For a discussion of several important cases in medical imaging where the singular value
  decomposition has been determined the reader can see \cite{Da,Ka,Kat,Lou,Maa,Maa1,Quell} and their references. It remains as a challenge to find commuting differential operators in most of these cases. For a case where this has been found in geophysical
  applications see \cite{GLP,SD,SDW}.

\bigskip

   The need to compute the singular value decomposition is, at times, of paramount importance.
   A good reconstruction algorithm should {\bf ONLY} try to find
   the projection of the object onto the span of the singular
   functions going with eigenvalues that are sensibly away from zero.

   This applies in particular
   to iterative algorithms such as those used in phase determination, see
   \cite{GS}. The same is true in the limited angle tomography problem.
  If this linear span is not
   satisfactory for the spatial resolution that one wants to achieve,
   one needs to measure over a larger range of angles, i.e. increase $\phi$, and compute
   the new singular
    value decomposition from scratch.

\bigskip

  In numerical simulations, where the phantom is known,
   the error (in well chosen examples) is seen to decrease monotonically as the number of iterations increases, but eventually the error starts increasing
erratically with the number of iterations. A good stopping criterium for deciding when to stop iterating is given by a detailed knowledge of the spectral
properties of the integral operator in question.

\bigskip

These integral operators with kernel given as in \eqref{kernel} have simple spectrum that rapidly accumulate at a point, making the effective computation of their eigenfunctions a very tricky problem. This problem is handled in \cite{SP,LP1,LP2,S,SV} very effectively by the miraculous existence of a commuting
differential operator: not only do these operators have the same eigenfunctions but the differential one has a very spread out spectrum yielding a numerically
stable way to obtain the desired singular functions. For a fuller discussion  of many important numerical aspects see \cite{ORX}.

\bigskip

It is appropriate to point out that the existence of these commuting differential operators for the corresponding integral operators  is a very exceptional situation. It plays an important role in discussions in Random Matrix theory, see \cite{Mehta} as well as for the Bessel and Airy cases, see \cite{TW1,TW2,TW3}.

\bigskip

The search for other situations where this (albeit exceptional) miracle
 holds is the motivation of this paper.

\bigskip

We are now in a position to state the new results in this paper.

We describe a method to extend the results of D. Slepian, \cite{S}, to families of differential operators that are deformations of the Bessel cases in \eqref{Bessel}. In the very short note \cite{Grunbaum1996} the first two examples to be discussed in Section $5$ were displayed without any indication of the method used
to produce them. These examples depend on one deformation paramenter, whereas the new examples presented in this paper depend on two parameters. All examples, new and old ones, are derived ab-initio in Section $5$.









We close this introduction by noticing that the presence of a deformation parameter in the operators \eqref{firstDarboux}, \eqref{secondDarboux}, \eqref{thirdDarboux} and \eqref{fourthDarboux}  for which we extend the results in \cite{S} is the reason for our interest in the KdV equation. The potentials $V(x)$ will be seen to evolve in time by an evolution equation very much related to the Korteweg-de Vries equation, as discussed in Section $4$. Before dealing with this point  we will see in Section $2$ how "free parameters" enter naturally in very basic constructions related to a general Schroedinger operator.

\section{The Darboux process}

We recall the ``Darboux process'', namely one that produces out of an initial second order (Schroedinger) operator

\begin{equation}
\label{Schroe}
L = -D_x^2 + V(x)
\end{equation}

\noindent
for which the eigenfunctions $\psi(x,z)$ are known, a new one parameter family of operators ${\tilde L}(t)$ whose eigenfunctions can be written in terms of those of $L$.

We can express $L$ in factorized form
\[
L = \left( -D_x - \frac {\phi'(x)}{\phi(x)} \right)\left( D_x - \frac {\phi'(x)}{\phi(x)}\right)
\]
where $\phi(x)$ is any eigenfunction of $L$ with zero eigenvalue and $\frac {\phi'(x)}{\phi(x)} = \partial_x(\log \phi)$. Since only the ratio $\frac {\phi'}{\phi}$ enters here one has
\[
\phi(x) = \phi^{(1)}(x) + t\phi^{(2)}(x)
\]
where $\phi^{(1)}(x)$, $\phi^{(2)}(x)$ form a basis of the two-dimensional space of eigenfunctions of $L$ with $0$ eigenvalue. This generic eigenfunction $\phi(x)$ depends now on the free deformation parameter $t$ and we will write it from now on as
\[
\phi(x,t).
\]
The operator ${\tilde L}$, i.e., the family of operators produced by the Darboux process, denoted by  ${\tilde L}(t)$, is now given by

\begin{equation}
{\tilde L}(t) \equiv \left( D_x - \frac {\phi'(x,t)}{\phi(x,t)}\right) \left( -D_x - \frac {\phi'(x,t)}{\phi(x,t)}\right)
\end{equation}

\noindent
and it is easy to see that

\begin{equation}
	\label{Crum}
{\tilde L}(t) = L - 2\partial_x^2 \log \phi(x,t).
\end{equation}

\noindent
If $\psi(x,z)$ is any solution of
\[
L\psi(x,z) = z^2\psi(x,z)
\]
then
\[
{\tilde L}(t)\left( D_x - \frac {\phi'(x,t)}{\phi(x,t)}\right) \psi(x,z) = z^2\left( D_x - \frac {\phi'(x,t)}{\phi(x,t)}\right) \psi(x,z),
\]
i.e., the eigenfunctions of ${\tilde L}(t)$ are given by
\begin{equation}
\label{eigCrum}
\left( D_x - \frac {\phi'(x,t)}{\phi(x,t)}\right) \psi(x,z).
\end{equation}

\bigskip

\noindent
We will make repeated use of the method in this section later in the paper.

\bigskip

For later use, see Section $5$, we note that if we define a function $\theta(x)$ by

$$V(x) = -\frac {1}{4x^2} - 2 \partial_x^2 \log \theta(x)$$

\noindent
and we put $$\phi(x,t)= \frac{{\tilde \theta(x,t)}}{\theta(x)}$$
we can express the operator in \eqref{Crum} above as

$${\tilde L}(t) = -D_x^2 -\frac{1}{4 x^2} - 2\partial_x^2 \log {\tilde \theta(x,t)}$$ 
and the eigenfunctions of this operator are given in terms of those of $L$, namely  $\psi(x,z)$, by the expression

$$ (1/z)\left(D_x - \partial_x \log \frac{{\tilde \theta(x,t)}}{\theta(x)}\right) \psi(x,z). $$

\bigskip

\noindent
This expression, with the factor $\frac{1}{z}$ included here for convenience will be useful in Section 5.

\section{The bispectral problem}
\label{sec3}

My interest in extending the results of the Bell Labs group mentioned earlier was the main motivation behind the problem posed and solved in \cite{DG86}, see the comments in page $178$ of that paper. That problem took a life of its own and the present paper is an attempt to return to this open question motivated (as mentioned earlier) by limited angle X-ray tomography.

For the benefit of the reader, we reproduce a few results from \cite{DG86} that will play an important role here. The sequence of functions of $x$

 \bigskip
 
\[
\begin{aligned}
\theta_0 =& 1 ,\\
\theta_1 =& x^{1/2} ,\\
\theta_2 =& x^2 + t_1 ,\\
\theta_3 =& \frac {3}{4} x^{9/2}+t_2 x^{1/2} ,\\
\theta_4 =& \frac {15}{32} x^8 + \frac {15t_2}{4} x^4 + t_3x^2 - \frac {5}{2} t_2^2 ,\\
\theta_5 =& \frac {525x^{25/2}}{2048} + \frac{35t_3x^{13/2}}{8} + \frac {3t_4x^{9/2}}{4} - \frac {7t_3^2x^{1/2}}{3} ,\\ 
\theta_6 =& \frac {33075}{262144} x^{18} + \frac {19845 t_3 }{2048} x^{12} + \frac {945 t_4 }{256} x^{10}  + \frac {15 t_5}{32} x^8 - \frac {2205 t^2_3}{32} x^6 - \frac {63 t_3 t_4 }{4} x^4
\\& + \left(t_3 t_5  - \frac {9 t_4^2}{5} \right)x^2 - \frac {49 t_3 ^3}{2} ,\\
\end{aligned}
\]
were obtained in \cite{DG86} by using the recursion
\[
\theta'_{k+1} \theta_{k-1} - \theta_{k+1}\theta'_{k-1} = (2k-1)\theta_k^2.
\] 
It is clear that at each step we can choose a new integration constant $t_i$. These give the deformation parameters mentioned earlier. The examples discussed in Section $5$ will only require the functions $\theta_2,\theta_3,\theta_4,\theta_5$ but we include $\theta_6$ here to make the dependence of these functions on the
free parameters more evident, see \eqref{dependence} and \eqref{dependence1}, below.

\bigskip

The functions above are shown in \cite{DG86} to give rise to one half of the solutions of the ``bispectral problem'', i.e.\ they allow one to define

\begin{equation}
\label{potentials}	
V_k = -\frac {1}{4x^2} - 2 \partial_x^2 \log \theta_k
\end{equation}

\noindent
such that the eigenfunctions $\phi(x,z)$ of
\[
L = -D_x^2 + V_k
\]
satisfy not only
\begin{equation}
\label{eq1}
L\phi(x,z) = z \phi(x,z)
\end{equation}
but also
\begin{equation}
\label{eq2}
B(z,\partial_{z})\phi(x,z) = \Theta(x)\phi(x,z)
\end{equation}
for some differential operator $B$.

\bigskip

In \cite{DG86} one proves that the potentials $V(x)$ in $L = -D_x^2 + V$ need to be rational functions for this very exceptional bispectral property to hold. The functions $\theta_k$ above meet this requirement if, as explained in \cite{DG86}, at every other step of the recursion we set some of the earlier $t_i$ equal to zero. This has the effect that the dependence of the functions $\theta_k$ on the free parameters $t_i$ is
given as follows  
\begin{align}
\label{dependence}
\theta_{2n}&=\theta_{2n}(x;t_n,t_{n+1},....,t_{2n-1})\\
\intertext{and }
	\label{dependence1}
\theta_{2n-1}&=\theta_{2n-1}(x;t_n,t_{n+1},....,t_{2n-2}),
\end{align} 
as can be seen in the explicit examples displayed above.

\bigskip

It may be appropriate to note that the other half of the solutions of the bispectral
problem considered in \cite{DG86}-and related to the Korteweg-de Vries hierarchy-will play no role in this paper. Their corresponding theta functions are obtained from the same recursion relation as above with $\theta_1=x$. In that case there is no need to set some of the $t_i$ equal to zero to get a rational potential $V(x)$. For different but related results dealing with this KdV hierarchy, or more generally the Kadomtsev-Petviashvilii hierarchy of equations, usually referred to as the KP
family of evolution equations, one can see \cite{CY,CGYZ1,CGYZ2}. For deep results on rational solutions of the KdV and the KP hierarchies, see \cite{AMM77,Krichever}. For other work on commuting integral-differential operators without the presence of deformation parameters but in a matrix valued context see \cite{CG,GPZ1,GPZ2}.

\bigskip

For the potentials $V_k(x)$ that we will be concerned with here, it was observed
in \cite{ZM} that the relevant evolution equations are the so called ``master symmetries of
Korteweg-de Vries''. The potentials $V_k$ were given in \cite{DG86} but the 
role of the master symmetries was only uncovered in \cite{ZM}.
The theta functions going with the KdV hierarchy are given by the characters of certain representations of $GL(n,R)$. Finding a similar group theoretical interpretation for the theta functions given above remains a challenge. 

\bigskip

We close this section with the remark that the desire to extend the results of
\cite{LP1,S,SP,SV} motivated by the consideration of the limited-angle X-ray
tomography problem has taken us and other authors pretty far into what some people may call ``pure'' mathematics. The sharp
distinction between pure and applied mathematics is, at times, not a productive one. The notion of bispectrality-explained above-surfaces from time to time in unexpected areas. For a recent instance see \cite{Nek}, and references [47-49] in that paper.

\section{ The KdV hierarchy and its master symmetries}

\bigskip

The celebrated Korteweg-de Vries equation from water wave theory 
is given, as was mentioned in the introduction, by
$$u_t = u_{xxx}- 6 u u_x.$$

This is not the place to give
any details about this subject, and we just mention a few relevant facts. There
are many references that could be cited, but for our purposes it is enough to refer
the reader to \cite{ZM,FZ} and the references there.

It is known that the KdV partial differential equation is the second one in a hierarchy of which the first
equation is just translation, given by solving 
\begin{equation}
\label{trans}
u_t=u_x
\end{equation}
with initial data
$$u(x,0)=f(x)$$
and solution
$$u(x,t)=f(x+t).$$

It is useful to denote the partial derivative operator acting on the function 
$u$ above as follows
$$u_x=\mathbb{X}_0(u)$$
so that \eqref{trans} becomes
$$u_t=\mathbb{X}_0(u).$$

\noindent

The remaining non-linear equations in the hierarchy, namely $$u_t= \mathbb{X}_j (u),$$

\noindent
can be obtained by applying higher order powers
of the tensor $\mathbb{N}_u$ 
to the generator of translations
$u_x=\mathbb{X}_0(u)$,
more explicitly
$$\mathbb{X}_j(u)= {\mathbb{N}_u^j} \;\mathbb{X}_0 (u)= {\mathbb{N}_u^j} \; u_x,  \quad j=2,3,4,\dots$$

\noindent
where the tensor $\mathbb{N}_u$ is given by
\begin{equation}
	\label{Niuh}
{\mathbb{N}_u} =-\partial_x^2+ 4 u + 2 u_x \partial_x^{-1}.
\end{equation}

One word about the operator of integration, denoted here by $\partial_x^{-1}$.
The arbitrary constant is chosen so that the resulting function vanishes at plus infinity.

\bigskip

An important property of the KdV evolution is that it is given by a Hamiltonian in the appropriate phase space of functions, an observation of L. Faddeev and V. Zakharov. Moreover it is what is called a bi-Hamiltonian flow, a notion going back to F. Magri and independently I. Dorfman, I.M. Gelfand and L. Dikii.
For a very good guide to all of these developments and the work of many other
authors who have made 
very important contributions in this area the reader may consult \cite{FZ}.

The consequence of this is that there exists another hierarchy of evolution equations obtained in this case by applying powers of the tensor \eqref{Niuh} to the infinitesimal generator of dilations, namely
 $$\mathbb{\tau}_0(u)= \frac{1}{2}x u_x +u.$$
 
\bigskip

This other hierarchy has generators given by
$$ \mathbb{\tau}_j(u)= {{\mathbb{N}^j_u}} \mathbb{\tau}_0(u),$$

\noindent
where the first few explict examples are given by 
\[
\begin{aligned}
\tau_0(u) &= \frac{1}{2} x u_x +u ,\\	
\tau_1(u)& = -\frac{x}{2}(u_{xxx}-6uu_x) -2 u_{xx} + u_x \partial_x^{-1} u + 4 u^2, \\
	\tau_2(u) &=\frac{x}{2}(u_{xxxxx}-10 u u_{xxx} -18 u_x u_{xx} +24 u^2 u_x) + 3 u_{xxxx} - u_{xxx} \partial_x^{-1} u \\
&\quad -24 u u_{xx} -15 u_x^2+ u_x ( 4 u \partial_x^{-1} u + 2 \partial_x^{-1} \tau_1(u))+16 u^3,
\end{aligned}
\]

\noindent
and the new family of evolution equations takes the form 
$$u_t= \tau_j(u).$$

\bigskip

As pointed out in \cite{vanMoerbeke}, see page $166$, very few explict solutions are known for these evolution equations which are transversal to the more familiar KdV hierarchy. Here we record a few such solutions in terms of the potentials $V_k$  in \eqref{potentials}. This establishes a remarkable connection between operators for which we extend in Section $5$ the result of D. Slepian, \cite{S}, and the master symmetries of the KdV equation.

\bigskip

We have 
\begin{align*}
\tau_1(V_3)&= 0\\
\intertext{and} 
\tau_0(V_3)&= -2 t_2 \frac {\partial V_3}{\partial t_2}.
\end{align*}

We also have 
\begin{align*}
\tau_1(V_4) &= - 120 t_2 \frac{\partial V_4}{\partial t_3}\\ 
\intertext{as well as} 
\tau_1(V_5) &= - 420 t_3 \frac{\partial V_5}{\partial t_4}\\ 
\intertext{and} 
\tau_2(V_6) &= - 105840 t_3 \frac{\partial V_6}{\partial t_5}.
\end{align*}

While the vector fields

$$\mathbb{X}_j(u)$$
going with the KdV hierarchy all commute with each other, the new vector fields

$$ \mathbb{\tau}_j(u)$$
satisfy the commutation relations

$$[\mathbb{X}_j,\mathbb{\tau}_l] = -(j+\frac{1}{2}) \mathbb{X}_{l+j}$$

\noindent
and

$$[\mathbb{\tau}_j,\mathbb{\tau}_l]= (l-j) \mathbb{\tau}_{l+j}.$$

\section{Commuting differential operators}

We have recalled in Section 3  the results from \cite{DG86} to obtain a sequence of families of operators of the form

\[
L(t) = -D_x^2 + V_k(x) 
\]

\noindent
where $V_k(x)$ is given by \eqref{potentials} and the dependence on $$t=(t_1,t_2,t_3,....)$$ results from the dependence of the functions $\theta_k$ on these parameters, and such that we have solutions of the bispectral problem.

\bigskip

In the first two families the operators
$L_2(t_1)$ and $L_3(t_2)$
depend on one free parameter, in the next two families we have two free parameters, in the following pair we have three 
free parameters, etc.

\bigskip

The purpose of this section is to discuss in detail a method to study the first four examples, starting with $\theta_2$, of the list of the $\theta_i$ given in Section 3 and in each case address the ``time-and-band limiting'' problem considered in \cite{SP,LP1,LP2,S,SV}.
More explicitly, we demonstrate the existence of a commuting differential operator that provides an effective and numerically stable way to compute the eigenfunctions of the naturally appearing integral operator in each case. For a careful look at the numerical issues involved here, see \cite{ORX}.
In the first two cases these results, without any indication of the method used to obtain them were given in \cite{Grunbaum1996}.

\subsection{
First example.}

\bigskip

We start with the operator   
\begin{equation}
\label{startnu=1)}
	L = - D_x^2  -\frac {1}{4x^2} - 2 \partial_x^2 \log \theta_1.
\end{equation}
which can be expressed as

$$L = - D_x^2  -\frac {1}{4x^2} + \frac{1}{x^2}= -D_x^2 +\frac{3}{4 x^2}.$$
 
\bigskip

This is one of the situations \eqref{Bessel} ($\nu=1$) covered by D. Slepian's
result, see \cite{S}, with (regular) eigenfunction given by \eqref{eigenf}.

One application of the Darboux method of Section $2$ produces the family of operators
\begin{equation}
\label{firstDarboux}
L_2(t_1) = - D_x^2  -\frac {1}{4x^2} - 2 \partial_x^2 \log \theta_2.
\end{equation}

In fact, as one can see in \cite{DG86}, this is the reason for introducing the sequence of functions $\theta_k$ that were reproduced in Section $3$.

As we observed at the end of Section $2$,  the eigenfunctions of this operator
can be written in terms of those of $L$, namely $f_1(x,z)$.

The new eigenfunction is 
\begin{equation}
	\label{firsteigen}
\widetilde{f}_1(x,z) = (1/z)(D_x - \partial_x \log \frac{\theta_2}{\theta_1}) f_1(x,z),
\end{equation} 
where, as we recall, $f_1$ is given in terms of the Bessel function $J_1$ and
thus $\widetilde{f}_1$ is written in terms of $J_1$ and $J_2$. The reason for this is the relation

\begin{equation}
\label{bessel}
\frac {\partial J_1}{\partial x}= \frac{1}{x} J_1 - J_2.
\end{equation}

These eigenfunctions depend on the free parameter $t_1$ brought in by the dependence of $\theta_1$ on it.

The integral kernel in question

\begin{equation}
\label{newkernel}
K(z_1,z_2) \equiv \int_0^T \widetilde{f}_1(x,z_1)\widetilde{f}_1(x,z_2)dx
\end{equation}

\noindent
can be written out explicitly in terms of $f_1(T,z_1)$, $f_1(T,z_2)$, $f_2(T,z_1)$ and $f_2(T,z_2)$, namely we have
\begin{multline*}
	K(z_1,z_2)= \frac {z_1 f_1(T,z_1)f_2(T,z_2)-z_2 f_2(T,z_1)f_1(T,z_2)}{z_1^2-z_2^2}\\+ \frac{2 t_1 f_1(T,z_1)f_1(T,z_2)}{(t_1+T^2)T z_1 z_2}.
\end{multline*}

\bigskip

A word about the derivation of this expression:  we note the fact that if a  function $f(x,z)$ satisfies

$$L f(x,z)= (-D_x^2 + V(x)) f(x,z)= z^2 f(x,z)$$

\noindent
it is straightforward to see that the expression 

\begin{equation}
\label{Christ}
	\frac {f(x,z_1) {\frac{\partial f(x,z_2)}{\partial_x}} - {\frac{\partial f(x,z_1)}{\partial_x}} f(x,z_2)}{z_1^2-z_2^2} 
\end{equation}

\bigskip

\noindent
is an antiderivative for the product $f(x,z_1) f(x,z_2)$.

\bigskip

Using this observation, the kernel \eqref{newkernel} has the traditional Christoffel-Darboux expression that follows from \eqref{Christ}. From it one can obtain the version of the kernel given above,
 free of any derivatives. This expression is more useful to us.

\bigskip

The same procedure will be used
	for the kernels in the next few examples. In every case we are making  repeated use of the relations \eqref{creation1x} and \eqref{creation2x} given below and involving \eqref{eigenf}

\begin{equation}
\label{creation1x}
\frac {\partial f_\nu}{\partial x}= \frac{1+2\nu}{2x} f_\nu -z f_{\nu+1}
\end{equation}
and

\begin{equation}
\label{creation2x}
\frac {\partial f_\nu}{\partial x}= \frac{1-2\nu}{2x} f_\nu +z f_{\nu-1}.
\end{equation}

\bigskip

Using the fact that $f_\nu$ depends only on the product $x z$ we
also have

\begin{equation}
\label{creation1z}
\frac {\partial f_\nu}{\partial z}= \frac{1+2\nu}{2z} f_\nu -x f_{\nu+1}
\end{equation}
and

\begin{equation}
\label{creation2z}
\frac {\partial f_\nu}{\partial z}= \frac{1-2\nu}{2z} f_\nu +x f_{\nu-1}.
\end{equation}

The second set of equations \eqref{creation1z} and \eqref{creation2z} will be used later in this section to get rid of $z$ derivatives.

\bigskip

The operator with kernel \eqref{newkernel}
acts on $L^2([-G,G],dz)$.

\bigskip

We set out to look for an operator of the form 
\begin{equation}
\label{ooppe}
\mathfrak {op}(z,\partial_z) = \sum_{k=0}^2 \partial_z^k (z^2-G^2)^k a_k(z)\partial_z^k, 
\end{equation}

\noindent
with $a_k(z)$ an even Laurent polynomial of degree $4-2k$, to be determined, such that 

\begin{equation}
\label{ooppe1}
\mathfrak{op}(z_1,\partial_{z_1}) K(z_1,z_2) = \mathfrak{op}(z_2,\partial_{z_2}) K(z_1,z_2).
\end{equation}

\bigskip

More explicitly , we look for $a_k(z)$ of the form $a_2(z)=1$, 
 $a_1(z)= a z^2+ b +\frac{c}{z^2}$ and $a_0(z)=d z^4+ e z^2 +f + \frac{g}{z^2} +\frac{h}{z^4}$ so that \eqref{ooppe1} will be satisfied.

\bigskip

The inclusion of the factors $(z^2-G^2)^k$ in the coefficients of the differential operator $\mathfrak{op}(z,\partial_z)$ guarantees that the domain where commutativity holds can be taken as the set of smooth functions that are bounded at the endpoints $z=G$ and $z=-G$. This consideration is important starting with the classical examples in \cite{SP,LP1,S}. A careful discussion of the domain of the differential
operator even in the classical cases is -to the best of my knowledge- only found in \cite{K}.

\bigskip

At this point we have another chance to use the well known relations among the functions $f_\nu(x,z)$ introduced earlier, see \eqref{creation1z} and \eqref{creation2z} (and very related to properties of the Bessel functions).
Using these relations repeteadly we can express any partial derivative (of any order) of $f_1(x,z)$ and $f_2(x,z)$ with respect to $z$ as linear combinations of $f_1(x,z)$ and $f_2(x,z)$ with coefficients that are rational functions of $x$ and $z$.

At the end of the day the commutativity relation \eqref{ooppe1} we are trying to satisfy (by a proper choice
of the $a_k$) amounts to the vanishing of a linear combination of the four products $$f_1(T,z_1) f_1(T,z_2),$$ $$f_2(T,z_1) f_1(T,z_2),$$ $$f_1(T,z_1) f_2(T,z_2),$$ and $$f_2(T,z_1) f_2(T,z_2).$$
with coefficients which are polynomials in $z_1,z_2,t_1,T$.
 The simplest such coefficient is the one going with the product $f_2(T,z_1) f_2(T,z_2)$, and it is given by
$$32 z_1^5 z_2^5 (z_2^2-z_1^2) T^2 (T^2+t_1)(2 T^2 - a),$$ 
where as we recall $a_1(z)= a z^2+ b +\frac{c}{z^2}$.

\bigskip

From here we see that $a=2 T^2$, and with the assistance of symbolic computation we can get that the vanishing of each of the four coefficients alluded to above gives an essentially unique solution for $a_0(z)$ and $a_1(z)$ once we take $a_2(z)=1$.

\bigskip

One gets
\begin{equation}
\label{a1}
a_1(z)=2 T^2 z^2 + \frac{4 G^2 t_1+1} {2} +  \frac {15 G^2}{2 z^2}
\end{equation}
and
\begin{equation}
\label{a0}
a_0(z)= T^4 z^4+ \frac{(4 G^2 t_1+9) T^2 z^2}{2} - \frac{4 G^4 t_1-15 G^2}{8 z^2}-\frac{135 G^4}{16 z^4}. 
\end{equation}

It is probably more interesting to notice that $\mathfrak{op}$ can be written in terms of the operators $\mathbb{A_\nu}$, see \eqref{commut}, that  Slepian found in the Bessel case, namely 

\begin{equation}
\label{nice1}
\mathfrak{op}= {\mathbb{A}_2}^2- \frac{3}{2} \mathbb{A}_2 - \frac {11 G^2 T^2}{2} + 2 t_1 G^2 \mathbb{A}_0.
\end{equation}
Notice that only ${\mathbb{A}_\nu}$ with even $\nu$ enter in \eqref{nice1}.

\bigskip

These results are summarized as follows

\bigskip

\begin{theorem}

The integral operator with the kernel given by \eqref{newkernel} acting in 
 $L^2([-G,G],dz)$ and the differential operator
$\mathfrak {op}(z,\partial_z)$, see \eqref{ooppe}, with $\allowbreak{a_2(z) = 1}$, $a_1(z)$  and $a_0(z)$ given as in \eqref{a1}, \eqref{a0} commute with each other.
Moreover the operator \eqref{ooppe} is given by a polynomial \eqref{nice1} in the operators
${\mathbb{A}_\nu}$, see \eqref{commut}, with coefficients that are polynomials
in all the parameters involved, namely $G,T$ and $t_1$.

\end{theorem}
\bigskip

Notice that  the familiy of operators \eqref{firstDarboux} for which we have
extended D. Slepian's result, \cite{S}, is explicitly given by

\[
L_2(t_1) = -D_x^2 + \frac {15x^4 - 18t_1x^2 - t^2_1}{4x^6 + 8t_1x^4 + 4t^2_1x^2}
\] 
and it interpolates by letting $t_1$ range from $0$ to $\infty$ between 
the operators $L_0$ and $L_2$ considered by D. Slepian. Finally, notice that this family, which is obtained by applying the Darboux process of section $2$ to $L_1$ does not include this operator.

\bigskip

A natural question here and in the examples that follow is : how do you
come up with a proposal for the order of the commuting differential operator 
$\mathfrak {op}(z,\partial_z)$, in \eqref{ooppe}?
The answer is that in each one of these ``bispectral" situations, see Section $3$
 we adopt the order of the operator
$B(z,\partial_{z})$
 \eqref{eq2}, worked out in \cite{DG86}.

%

\subsection{Second example.}

\

%
\bigskip

Here we start with the operator $L_2$, see \eqref{Bessel}, and apply the Darboux process to it.
This happens to be the operator obtained in the example above when $t_1=0$. 

\bigskip

We obtain the family of operators 
\begin{equation}
\label{secondDarboux}
	L_3(t_2) = - D_x^2  -\frac {1}{4x^2} - 2 \partial_x^2 \log \theta_3.
\end{equation}

This family of operators, has eigenfunctions given by

\[
	\widetilde{f}_2(x,z) = (1/z)\left(D_x - \partial_x \log \frac{\theta_3}{\theta_2}\right) f_2(x,z),
\]
where, as we recall, $f_2$ is given in terms of the Bessel function $J_2$ and thus $\widetilde{f}_2$ is given in terms of $J_2$ and $J_3$. In the previous example the important functions were $J_1$ and $J_2$. In this and the examples in the next subsections this role is taken up by $J_2$ and $J_3$.

\bigskip

In the expression above
for $\widetilde{f}_2$ 
we have set $t_1=0$ and the eigenfunctions depend on the free parameter $t_2$ brought in by the function $\theta_3$ in one application
of the Darboux process.

\bigskip

The integral kernel in question

\begin{equation}
\label{newkernel2}
K(z_1,z_2) \equiv \int_0^T \widetilde{f}_2(x,z_1)\widetilde{f}_2(x,z_2)dx
\end{equation} 
can be written out explicitly in terms of $f_2(T,z_1)$ , $f_2(T,z_2)$, $f_3(T,z_1)$ and $f_3(T,z_2)$, namely we have

\begin{multline}
	K(z_1,z_2)= \frac {z_1 f_2(T,z_1)f_3(T,z_2)-z_2 f_3(T,z_1)f_2(T,z_2)}{z_1^2-z_2^2} \\- \frac{4 t_2 f_2(T,z_1)f_2(T,z_2)}{(t_2+T^4)T z_1 z_2}.
\end{multline}

We set out to look for an operator of the form

\begin{equation}
\label{ooppe2}
\mathfrak{op}(z,\partial_z) = \sum_{k=0}^3 \partial_z^k (z^2-G^2)^k a_k(z)\partial_z^k,
\end{equation} 
with $a_k(z)$ an even Laurent polynomial of degree $6-2k$ such that
$$\mathfrak{op}(z_1,\partial_{z_1}) K(z_1,z_2) = \mathfrak{op}(z_2,\partial_{z_2}) K(z_1,z_2).$$

By using the same strategy as in the first example
we can express all partial derivatives of $f_2(x,z)$ and $f_3(x,z)$ with respect to $z$ as linear combinations of $f_2(x,z)$ and $f_3(x,z)$ with coefficients that are rational functions of $x$ and $z$.

Finally we need to insure that the coefficients
of the four products 
\begin{align*}f_2(T,z_1) f_2(T,z_2)&,\\
f_3(T,z_1) f_2(T,z_2)&,\\
f_2(T,z_1) f_3(T,z_2)&,\\[-10pt]
\intertext{and \vskip-5pt} 
f_3(T,z_1) f_3(T,z_2)&,
\end{align*}
will vanish. The resulting set of equations for $a_k$ can be solved uniquely once we take $a_3(z)=1$.

\bigskip

Once again we can write $\mathfrak{op}$, see \eqref{ooppe2}, 
in terms of the operators $\mathbb{A_\nu}$ that  Slepian found in the Bessel case, namely

\begin{equation}
\label{nice2}
\mathfrak{op} = {\mathbb{A}_3}^3 - \frac{29}{4}{\mathbb{A}_3}^2 + \frac {195-256 G^2 T^2}{16} {\mathbb{A}_3} + \frac {435 G^2 T^2}{8}
+ 3 G^2 t_2 {\mathbb{A}_1}.
\end{equation} 
Notice that only ${\mathbb{A}_\nu}$ with odd $\nu$ enter in the expression \eqref{nice2}.

\bigskip

These results are summarized as follows

\bigskip

\begin{theorem}

The integral operator with kernel given by \eqref{newkernel2} acting in 
 $L^2([-G,G],dz)$ and the differential operator
$\mathfrak {op}(z,\partial_z)$, see \eqref{ooppe2}, with $a_3(z)=1$, $a_2(z), a_1(z)$ and $a_0(z)$ properly chosen commute with each other.
Moreover the operator \eqref{ooppe2} is given by a polynomial \eqref{nice2} in the operators
${\mathbb{A}_\nu}$, see \eqref{commut}, with coefficients that are polynomials
in all the parameters involved, namely $G,T$ and $t_2$.
\end{theorem}

Notice that  the familiy of operators \eqref{secondDarboux} for which we have now
extended D. Slepian's result, \cite{S}, is explicitly given by

\[
L_3(t_2) = -D_x^2 + \frac {35x^8 - 90t_2x^4 + 3t_2^2}{4x^{10} + 8t_2x^6 + 4t_2^2x^2}.
\] 
and it interpolates by letting $t_2$ range from $0$ to $\infty$ between 
the operators $L_1$ and $L_3$ considered by D. Slepian. Finally, notice that this family, which obtained by applying the Darboux process of section $2$ to $L_2$ does not include this operator.

\bigskip
%
%

\subsection{Third example.}

\

\bigskip
%

In this subsection we start with any of the operators in \eqref{secondDarboux} considered in the previous example and perform a step of the Darboux process to it, bringing in a new parameter $t_3$ besides $t_2$.

This results in the two parameter family of operators

\begin{equation}
\label{thirdDarboux}
L_4(t_2,t_3) = - D^2  -\frac {1}{4x^2} - 2 \partial_x^2 \log \theta_4.
\end{equation}

This family of operators, has eigenfunctons given by 
\[
\widetilde{f}_3(x,z) = (1/z)(D_x - \partial_x \log \frac{\theta_4}{\theta_3}) \widetilde{f}_2(x,z) 
\]
where $\widetilde{f}_2$ are the eigenfunctions of the previous example. Eventually they can be expressed in terms of the Bessel functions $J_2$ and $J_3$.

The integral kernel in question 
\begin{equation}
\label{newkernel3}
K(z_1,z_2) \equiv \int_0^T \widetilde{f}_3(x,z_1)\widetilde {f}_3(x,z_2)dx
\end{equation} 
can be written out explicitly in terms of $f_2(T,z_1)$ , $f_2(T,z_2)$, $f_3(T,z_1)$ and $f_3(T,z_2)$.

\bigskip
We set out to look for an operator of the form 
\begin{equation}
\label{ooppe3}
\mathfrak{op}(z,\partial_z) = \sum_{k=0}^5 \partial_z^k (z^2-G^2)^k a_k(z)\partial_z^k,
\end{equation}

with $a_k(z)$ an even Laurent polynomial of degree $10-2k$ such that

$$\mathfrak{op}(z_1,\partial_{z_1}) K(z_1,z_2) = \mathfrak{op}(z_2,\partial_{z_2}) K(z_1,z_2).$$

\bigskip

By now the strategy should be clear.
We can express all partial derivatives of $f_2(x,z)$ and $f_3(x,z)$ with respect to $z$ as linear combinations of $f_2(x,z)$ and $f_3(x,z)$ with coefficients that are rational functions of $x$ and $z$ and then try to insure the
vanishing of each coefficient going with the four products 

\begin{align*}
f_2(T,z_1) f_2(T,z_2),& \\f_3(T,z_1) f_2(T,z_2),& \\f_2(T,z_1) f_3(T,z_2),
\intertext{and}f_3(T,z_1) f_3(T,z_2).&
\end{align*}

\noindent
The vanishing of the individual coefficients determines the differential operator \eqref{ooppe3} once we set $a_5(z)=1$.

\bigskip

Once again we can write $\mathfrak{op}$ 
in terms of the operators $\mathbb{A_\nu}$ that  Slepian found in the Bessel case, namely 
\begin{equation}
\label{nice3}
\begin{aligned}
\mathfrak{op} = &\sum_{j=0}^5 w_j {\mathbb{A}_4}^j + t_2^2 \left( -\frac{320}{9} G^8 {\mathbb{A}_2} + \frac{80}{9} G^8 {\mathbb{A}_4}\right)   \\ 
&+ t_3 \left( \frac{ 16 G^6}{3} {\mathbb{A}_2}^2 - 8 G^6 {\mathbb{A}_2} - \frac {88 G^8 T^2}{3}\right) \\
& + t_2 \left(v_1 {\mathbb{A}_2}+v_2{\mathbb{A}_0^3}+ v_3{\mathbb{A}_4}+v_4{\mathbb{A}_0^2}+ v_5{\mathbb{A}_2^2}+ v_6{\mathbb{A}_2^3}+v_7 I \right), \\
\end{aligned}
\end{equation}
with the $w_i$ given by   
\begin{align*}
w_5&=1 ,\quad w_4=-{\frac{95}{4}},\quad w_3={-\frac{320\,G^2\,T^2-1549}{8}},\\ w_2&=\frac{22336\,G^2\,T^2-19703}{32}, \quad w_1 = {\frac{36864\,G^4\,T^4-971648\,G^2\,T^2+162645
	}{256}} ,\\ w_0&=-\frac {{84726\,G^4\,T^4-399805\,G^2\,T^2}}{64},
\end{align*}\\
and the $v_i$ given by
\begin{align*}
v_1&=-{\frac{5\,G^4\,\left(64\,G^2\,T^2-363\right)}{9
 }} , v_2=-{\frac{40\,G^4}{3}},
v_3=-{\frac{5\,G^4\,\left(  64\,G^2\,T^2+69\right)}{18}},\\
 v_4&=-{\frac{70\,G^4}{3}} ,\quad
 v_5=-{\frac{340\,G^4}{3}},\quad
   v_6={\frac{80\,G^4}{3}},\quad
 v_7=-{\frac{305\,G^6\,T^2}{3}}.
\end{align*}

\bigskip

These results are summarized as follows

\bigskip

\begin{theorem}

The integral operator with kernel given by \eqref{newkernel3} acting in 
 $L^2([-G,G],dz)$ and the differential operator
$\mathfrak {op}(z,\partial_z)$, see \eqref{ooppe3}, with $a_5(z)=1$, $a_4(z), a_3(z), a_2(z), a_1(z), a_0(z)$ properly chosen commute with each other.
Moreover the operator \eqref{ooppe3} is given by a polynomial \eqref{nice3} in the operators
${\mathbb{A}_\nu}$, see \eqref{commut}, with coefficients that are polynomials
in all the parameters involved, namely $G,T$ and $t_2,t_3$.

\end{theorem}

\bigskip

Notice that only ${\mathbb{A}_\nu}$ with even $\nu$ enter in \eqref{nice3}.

%
%
%
%
\bigskip

\subsection{
Fourth example.}

\
\bigskip
%
%
%

Here we start with the operator \eqref{thirdDarboux} but we set the parameter
$t_2$ equal to $0$ and apply one step pf the Darboux process to get a family
of operators depending on the old free parameter $t_3$ and a new parameter $t_4$.

This results in the family of operators 

\begin{equation}
\label{fourthDarboux}
L_4(t_3,t_4) = - D^2  -\frac {1}{4x^2} - 2 \partial_x^2 \log \theta_5.
\end{equation}

\bigskip

\noindent
This operator, has eigenfunctions given by 
\[
\widetilde{f}_4(x,z) = (1/z)(D - \partial_x \log \frac{\theta_5}{\theta_4}) \widetilde{f}_3(x,z),
\]
where $\widetilde{f}_3$ are the eigenfunctions of the previous example, with the understanding that $t_2=0$. Eventually they can be expressed in terms of the Bessel functions $J_2$ and $J_3$. 

\bigskip

The integral kernel in question
\begin{equation}
\label{newkernel4}
K(z_1,z_2) \equiv \int_0^T \widetilde{f}_4(x,z_1)\widetilde {f}_4(x,z_2)dx
\end{equation} 
can be written out explicitly in terms of $f_2(T,z_1)$ , $f_2(T,z_2)$, $f_3(T,z_1)$ and $f_3(T,z_2)$.

\bigskip
We set out to look for an operator of the form
\begin{equation}
\label{ooppe4}
\mathfrak{op}(z,\partial_z) = \sum_{k=0}^7 \partial_z^k (z^2-G^2)^k a_k(z)\partial_z^k,
\end{equation} 
with $a_k(z)$ an even Laurent polynomial of degree $14-2k$ such that
$$\mathfrak{op}(z_1,\partial_{z_1}) K(z_1,z_2) = \mathfrak{op}(z_2,\partial_{z_2}) K(z_1,z_2).$$

\bigskip

%
%


We follow the same strategy as above and determine the differential operator.

\bigskip

As in the previous examples we can write $\mathfrak{op}$ 
in terms of the operators $\mathbb{A_\nu}$ that  Slepian found in the Bessel case, namely
\begin{equation}
\label{nice4}
\begin{aligned}
\mathfrak{op}& = \sum_{j=0}^7 w_j {\mathbb{A}_5}^j  + t_3^2 ( -\frac{7168}{225 } G^{12} (3{\mathbb{A}_3} -  {\mathbb{A}_5}))   + t_4 \sum_{j=0}^3 v_j {\mathbb{A}_3}^j \\
\,\,&{	\hskip-7pt}+ t_3 (u_1 {\mathbb{A}_3}^4+u_2{\mathbb{A}_1^4}+ u_3{\mathbb{A}_3}^3+u_4{\mathbb{A}_1^3}+ u_5{\mathbb{A}_3^2}+ u_6{\mathbb{A}_1^2}+u_7{\mathbb{A}_3}+ u_7{\mathbb{A}_1}+u_7 I ) .
\end{aligned}
\end{equation}
where the explicit expression for the coefficients $w_j,v_j,u_j$ is rather unilluminating and will not be displayed.

\bigskip

These results are summarized as follows

\bigskip

\begin{theorem}
The integral operator with kernel  given by  \eqref{newkernel4} acting in 
 $L^2([-G,G],dz)$ and the differential operator
$\mathfrak {op}(z,\partial_z)$, see \eqref{ooppe4}, with $a_7(z)=1$,  $a_6(z), a_5(z), a_4(z), a_3(z), a_2(z), a_1(z), a_0(z)$ properly chosen commute with each other.
Moreover the operator \eqref{ooppe4} is given by a polynomial \eqref{nice4} in the operators
${\mathbb{A}_\nu}$, see \eqref{commut}, with coefficients that are polynomials
in all the parameters involved, namely $G,T$ and $t_3,t_4$.
\end{theorem}

\bigskip

Notice that only ${\mathbb{A}_\nu}$ with odd $\nu$ enter in \eqref{nice3}.















\bigskip



\section{Final comments}

\bigskip

The results in \cite{CY,CGYZ1,CGYZ2} deal with the large class of situations related to the KP hierarchy, we have only dealt here with the master symmetries
of the KdV evolution equations. These are related to the Schroedinger (second order) differential operator. The paper \cite{Grunbaum1986} considers solutions of the bispectral problem when this second order differential operator is replaced by a third order one. The  possible existence of commuting pairs of integral and differential operators in this case is -to the best of knowledge- largely unexplored territory. For interesting examples and theoretical tools that could be applied in this case, see \cite{CY}.

\bigskip

The problem considered here is an extension in the case of higher dimensional
Euclidean spaces of the work in \cite{S}. In the paper \cite{GLP}
one finds an excursion into the non-commutative situation that arises when Euclidean space is replaced by spheres. That study was motivated purely by mathematical reasons, and yet several years later, people working in geophysics found these results to have practical use, see \cite{SD,SDW}.

\end{document}